\documentstyle[11pt,adassconf]{article}  

\begin{document}   

%

\paperID{O3-03}


\title{Exploiting VSIPL and OpenMP for Parallel Image Processing}

%

\author{Jeremy Kepner\altaffilmark{1}}
\affil{Massachusetts Institute of Technology Lincoln Laboratory
       Lexington, MA 02420, Email: kepner@ll.mit.edu}

\altaffiltext{1}{This work is sponsored by DARPA, under Air Force Contract
F19628-00-C-0002.  Opinions, interpretations, conclusions and
recommendations are those of the author and are not necessarily
endorsed by the United States Air Force.}



\contact{Jeremy Kepner}
\email{KEPNER@LL.MIT.EDU}

%
%

\paindex{Kepner, J.}




\begin{abstract}          

  VSIPL and OpenMP are two open standards for portable high
performance computing.  VSIPL delivers optimized single processor
performance while OpenMP provides a low overhead mechanism for executing
thread based parallelism on shared memory systems. Image processing is
one of the main areas where VSIPL and OpenMP can make a large impact. 
Currently, a large fraction of image processing applications are written
in the Interpreted Data Language (IDL) environment.  The aim of this
work is to demonstrate that the performance benefits of these new
standards can be brought to image processing community in a high level
manner that is transparent to users. To this end, this talk presents a
fast, FFT based algorithm for performing image convolutions.  This
algorithm has been implemented within the IDL environment using VSIPL
(for optimized single processor performance) with added OpenMP
directives (for parallelism). This work demonstrates that good parallel
speedups are attainable using standards and can be integrated seamlessly
into existing user environments.
\end{abstract}


\section{Introduction}

  The Vector, Signal and Image Processing Library (VSIPL) [1] is an
open standard C language Application Programmer Interface (API) that
allows portable and optimized single processor programs.   OpenMP [2]
is an open standard C/Fortran API that allows portable thread based
parallelism on shared memory computers. Both of these standards have
enormous potential to allow users to realize the goal of portable
applications that are both parallel and optimized.

  Exploiting these new open standards requires integrating them into
existing applications as well as using them in new efforts.  Image
processing is one of the key areas where VSIPL and OpenMP can make a
large impact.  Currently, a large fraction of image processing
applications are written in the Interpreted Data Language (IDL)
environment [3].  The goal of this work is to show that it is possible to
bring the performance benefits of these new standards to the image
processing community in a high level manner that is transparent to
users.

\section{Approach}

\begin{figure}
  \epsscale{.80}
  \label{basic_2d_filtering}
  \plotone{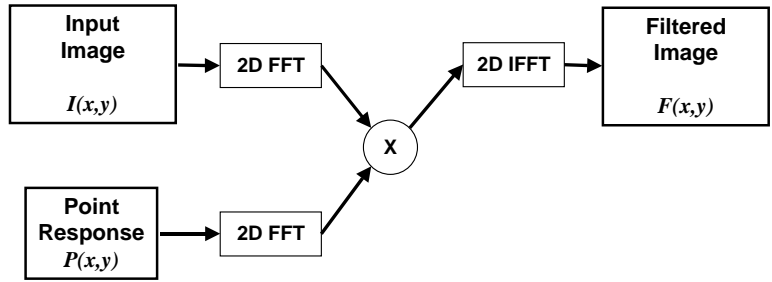}
  \caption{{\bf Basic 2D Filtering.}  FFT implementation of
    2D filtering which performs the mathematical operation:
    $F(x,y) = \int \int P(x',y') I(x - x',y - y') dx' dy'$
  }
\end{figure}

  Wide area 2D convolution is a staple of digital image processing (see
Figure~\ref{basic_2d_filtering}). The advent of large format CCDs makes
it possible to literally ``pave'' with silicon the focal plane of an
optical sensor. Processing of the large images obtained from these
systems is complicated by the non-uniform Point Response Function (PRF)
that is common in wide field of view instruments. This paper presents a
fast, FFT based algorithm for convolving such images.  This algorithm
has been transparently implemented within IDL environment using VSIPL
(for optimized single processor performance) with added OpenMP
directives (for parallelism).

\begin{figure}
  \epsscale{.80}
  \label{wide_field_filtering}
  \plotone{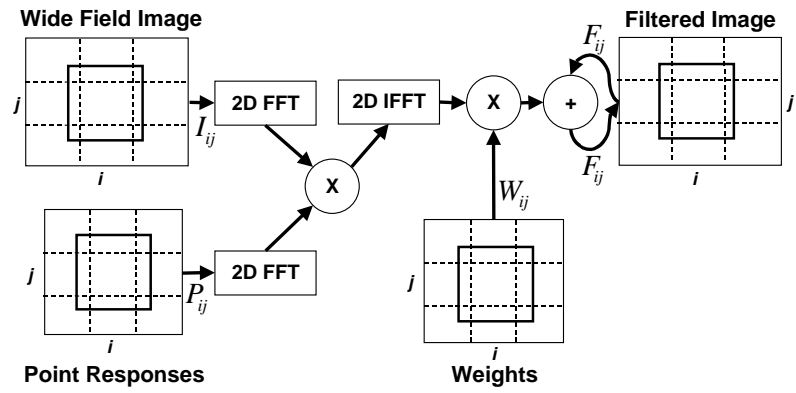}
  \caption{{\bf Wide Field Filtering.}  FFT implementation of
    2D filtering for wide field imaging with multiple point response
    functions.  Each portion of image is filtered separately and
    then recombined using the appropriate weights.  The equivalent
    mathematical operation is:
    $F_{ij}(x,y) = W_{ij}(x,y) \int \int P_{ij}(x',y') I_{ij}(x - x',y - y') dx' dy'$
  }
\end{figure}

  The inputs of image convolution with variable PRFs consists of a
source image, a set of PRF images, and a grid which locates the center
of each PRF on the source image.  The output image is the convolution of
the input image with each PRF linearly weighted by its distance from its
grid center. The computational basis of this convolution are 2D overlap
and add FFTs with interpolation (see Figure~\ref{wide_field_filtering}).
Today, typical images sizes are in the millions (2K x 2K) to billions
(40K x 40K) of pixels.  A single PRF is typically thousands of pixels
(100 x 100) pixels, but can be as small 10 x 10 or as large as the
entire image.  Over a single image a PRF will be sampled as few as once
but as many as hundreds of times depending on the optical system.

\begin{figure}
  \epsscale{.80}
  \label{software_layers}
  \plotone{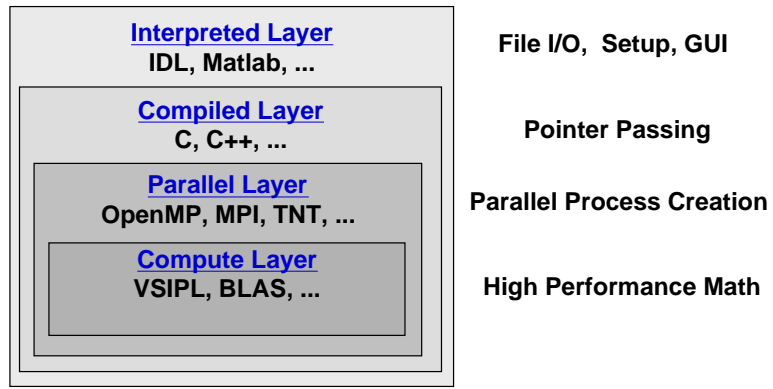}
  \caption{{\bf Layered Software Architecture.}  The user interacts
    with the top layer which provides high level abstractions for
    high productivity.  Lower layers provide performance via parallel
    processing and high performance kernels.
  }
\end{figure}

  There are many opportunities for parallelism in this algorithm. The
simplest is to convolve each PRF separately on a different processor and
then combine all the results on a single processor.  This approach works
well with VSIPL, OpenMP and IDL (see Figure~\ref{software_layers}.  At
the top level a user passes the inputs into an IDL routine which passes
pointers to an external C function.  Within the C function OpenMP forks
off multiple threads.  Each thread executes its convolution using VSIPL
functions.  The OpenMP threads are then rejoined and the results are
added.  Finally a pointer to the output image is returned to the IDL
environment in the same manner as any other IDL routine.

\section{Results}

\begin{figure}
  \epsscale{.80}
  \label{parallel_performance}
  \plotone{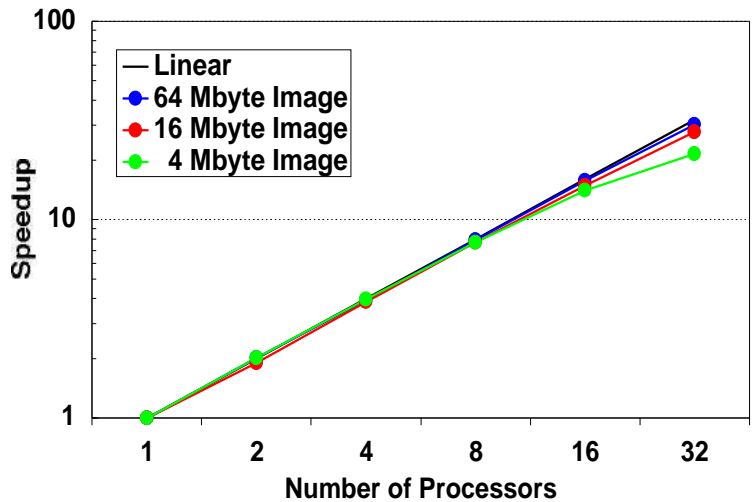}
  \caption{{\bf Parallel Performance.}  Measured speedups of
     wide field 2D filtering application on an shared memory
     parallel system (SGI Origin2000).  Results indicate linear
     speedups are achievable using open software standards
     underneath high level programming languages.
  }
\end{figure}

  This algorithm was implemented on an SGI Origin 2000 at Boston
University [4].  This machine consists of 64 300 MHz MIPS 10000
processors with an aggregate memory of 16 GBytes.  IDL version 5.3 from
Research Systems, Inc. was used along with SGI's native OpenMP compiler
(version 7.3.1) and the TASP VSIPL implementation. Implementing the
components of the system was the same as if each were done separately. 
Integrating the pieces (IDL/OpenMP/VSIPL) was done quickly, although
care must be taken to use the latest versions of the compilers and
libraries.  Once implemented the software can be quickly ported via
Makefile modifications to any system that has IDL, OpenMP, and VSIPL
(currently these are SGI, HP, Sun, IBM, and Red Hat Linux). We have
conducted a variety of experiments which show linear speedups using
different numbers of processors and different image sizes (see
Figure~\ref{parallel_performance}).  Thus, it possible to achieve
good performance using open standards underneath existing high level
languages.

\end{document}